# A Multifunctional Capacitive Sensing Platform for Wireless Vascular and Heart Monitoring

Parviz Zolfaghari[1,*], Beril Yagmur Koca[1], Taher Abbasiasl[1], Hakan Urey[1,2], and Hadi Mirzajani[1,*]

[1] Department of Electrical and Electronics Engineering, Koç University, 34450 Istanbul, Türkiye
[2] Koç University Research Center for Translational Medicine (KUTTAM), 34450 Istanbul, Türkiye

* pzolfaghari@ku.edu.tr,
* hmirzajani@ku.edu.tr

**Abstract**

We present a multifunctional, antenna-integrated capacitive sensing (MAiCaS) platform for passive, wireless, and real-time cardiovascular monitoring. Unlike conventional systems that require separate sensors and wireless modules, our device unifies sensing, telemetry, and mechanical functionality into a compact and scalable design by exploiting the parasitic capacitance of an inductive antenna as a strain-sensitive element. The sensor is fabricated using a cleanroom-free, single-step UV laser patterning process on a flexible PDMS substrate, significantly reducing manufacturing complexity and enabling high reproducibility. The MAiCaS is suitable for three different applications: as a sensor for epicardial strain measurement, a stent as a sensor, and a vascular graft sensor. We demonstrate MAiCaS's versatility by validating its wireless resonance-based response to strain, pressure, and deformation across unrolled and rolled forms. In vitro experiments demonstrated consistent resonance frequency shifts under physiological conditions, with stable performance observed on skin, in PBS, human serum, and simulated vascular environments. Repeatability and aging tests confirmed its long-term reliability and elasticity under cyclic loading. Calibration curves revealed high sensitivity across all configurations, with wireless interrogation achieved through $S_{11}$ parameter measurements and resonance frequency shift as the output metric. The sensitivity of the device was measured to be 2.9 MHz per 1% strain, 0.43 MHz/mmHg, and 309.6kHz/μm for applications of epicardial patch, graft, and stent integrated sensor, respectively. Furthermore, the operation of MAiCaS was evaluated in a human experiment using the unrolled format. This monolithic sensor architecture provides a scalable and cost-effective solution for battery-free monitoring of vascular dynamics, with strong potential for remote diagnostics, post-surgical follow-up, and continuous cardiovascular health management.

## 1. Introduction

Cardiovascular diseases (CVDs) are the leading cause of death globally [1]. In 2019, CVD accounted for approximately 17.9 million deaths (~32% of global mortality), rising to 20.5 million in 2021 (~33%) [2, 3]. In the European Union, circulatory diseases resulted in around 1.71 million deaths in 2021, representing 32.4% of total mortality [4]. In 2023, in the United States, 919,032 people died from cardiovascular disease, which is the equivalent of about one in every three deaths [5]. These life-threatening conditions often stem from vascular pathologies, such as arterial stenosis, aneurysmal dilation, and loss of elastic compliance, which contribute to catastrophic events, including myocardial infarction, ischemic stroke, aortic rupture, and sudden cardiac death [6].

To reduce mortality and improve long-term outcomes, patients at risk of serious cardiovascular events frequently undergo surgical or catheter-based interventions. These include deploying vascular stents to restore or maintain vessel patency or implanting vascular grafts to bypass diseased or occluded vascular segments [7]. While these implants provide essential mechanical support and restore physiological blood flow, they lack embedded diagnostic functionality for continuous post-operative monitoring. This absence of real-time feedback hinders early detection of complications like in-stent restenosis [8], which can still occur in approximately 5–10% of modern drug-eluting stent cases, and graft failure, with saphenous vein graft patency rates descending to 50–60% at 5–10 years post-implantation [9].

To bridge this gap, a new class of implantable smart devices, including sensorized stents and vascular grafts, has emerged. These systems combine traditional mechanical support with embedded physiological sensing and wireless data transmission. Typically, they integrate capacitive or piezoresistive strain or pressure sensors capable of detecting local tissue deformation or flow changes, paired with wireless telemetry systems based on inductive coupling or resonant inductive-capacitive (LC) circuits for data readout [10-12].

Several self-reporting stents have been proposed for real-time monitoring of hemodynamics and restenosis. Herbert et al. developed a wireless vascular system with a multimaterial inductive stent and printed pressure sensors [11]. Oyunbaatar et al. introduced a self-rollable polymer stent with a monolithically integrated LC pressure sensor, enabling wireless tracking in various media [2]. Lee et al. demonstrated a battery-free soft stent with nanomembrane strain sensors for restenosis monitoring via inductive telemetry [13]. A follow-up design by Oyunbaatar et al. integrated dual capacitive sensors and an enlarged inductor coil, achieving ~0.15 MHz/mmHg sensitivity [14]. Other advances include radar-readable metallic stents [15], extra-arterial LC sensors with 8.53 kHz/mmHg sensitivity [16], and an ANSYS-optimized PDMS-based capacitive sensor having a sensitivity of 10.68 fF/mmHg [17]. Recent advances in smart vascular grafts (SVGs) have enabled continuous hemodynamic monitoring. Ma et al. developed a piezoelectric vascular graft (PVG) using Polyvinylidene Fluoride (PVDF) nanofibers and AgNW electrodes encapsulated in polycaprolactone (PCL), achieving 11 mV/kPa sensitivity [18]. In a follow-up study, Ma et al. introduced an SVG with embedded porous graphene flow sensors in a biopolymer matrix, offering 0.0034% strain detection and remote readout in a rabbit model [19]. Hacohen et al. proposed a flexible, batteryless RFID-enabled flow sensor with a split double-helix antenna and piezoresistive nanocomposite sensor, capable of measuring pulsatile flow changes as low as 10 mL/min [20]. These studies demonstrate diverse strategies for real-time vascular monitoring.

While these devices are highly effective for monitoring vascular conditions, they face significant fabrication and integration challenges. Typically, a miniaturized capacitive sensor must be co-designed with a separate inductive antenna, followed by complex system integration. This increases design complexity, alignment sensitivity, and overall makes the fabrication and integration challenging. Furthermore, most existing platforms are engineered for a single function, either as a stent or a graft, and lack structural or functional adaptability for other applications.

To overcome the fabrication and integration challenges faced by existing designs and enhance the adaptability of the device for different applications, we present MAiCaS, a multifunctional, antenna-integrated capacitive sensing platform that unifies mechanical support, physiological sensing, and wireless communication within a single, compact, and scalable structure. Rather than relying on discrete sensors and antennas, our design re-engineers the antenna path itself to serve as the sensing element. Specifically, we exploit the parasitic capacitance of an inductive antenna line, enhanced through embedded interdigitated electrodes (IDE), to form a capacitive strain sensor. This dual-purpose conductive path enables both wireless telemetry and mechanical sensing, eliminating the need for separate sensor fabrication, antenna coupling, and interface electronics. Importantly, our device avoids the complex, multi-step, cleanroom-dependent processes common to previous systems. Instead, it is fabricated in a single, cleanroom-free laser patterning step on a flexible substrate, allowing for rapid prototyping, reproducible results, and scalable production. Owing to its monolithic and flexible architecture, the platform can be adapted into three distinct configurations (Figure 1a, b, and c): (i) smart cardiac patch: attached to heart provide post-surgery continuous monitoring to track biomechanical compliance in conditions like hypertension [21-24], arrythmia [25-28], and aneurysms and cardiomyopathies [29, 30], (ii) smart vascular graft: functions as both a load-bearing conduit and wireless diagnostic unit, for aortic repair, congenital defect correction, or post-surgical monitoring [11, 31, 32], and (iii) smart stent: enabling real-time sensing of restenosis or localized pressure changes [15, 33]. By integrating sensing and wireless communication roles into a single antenna structure, our platform represents an advancement in implantable sensing devices for cardiovascular applications. It simplifies fabrication, eliminates system integration complexities, and enables multifunctional use. After implementation in the specific body organ (according to the specific application of the device), continuous wireless monitoring is possible through a hand-held frequency reading device, where the results can be wirelessly transmitted to local hospitals or emergency departments for real-time monitoring and intervention.

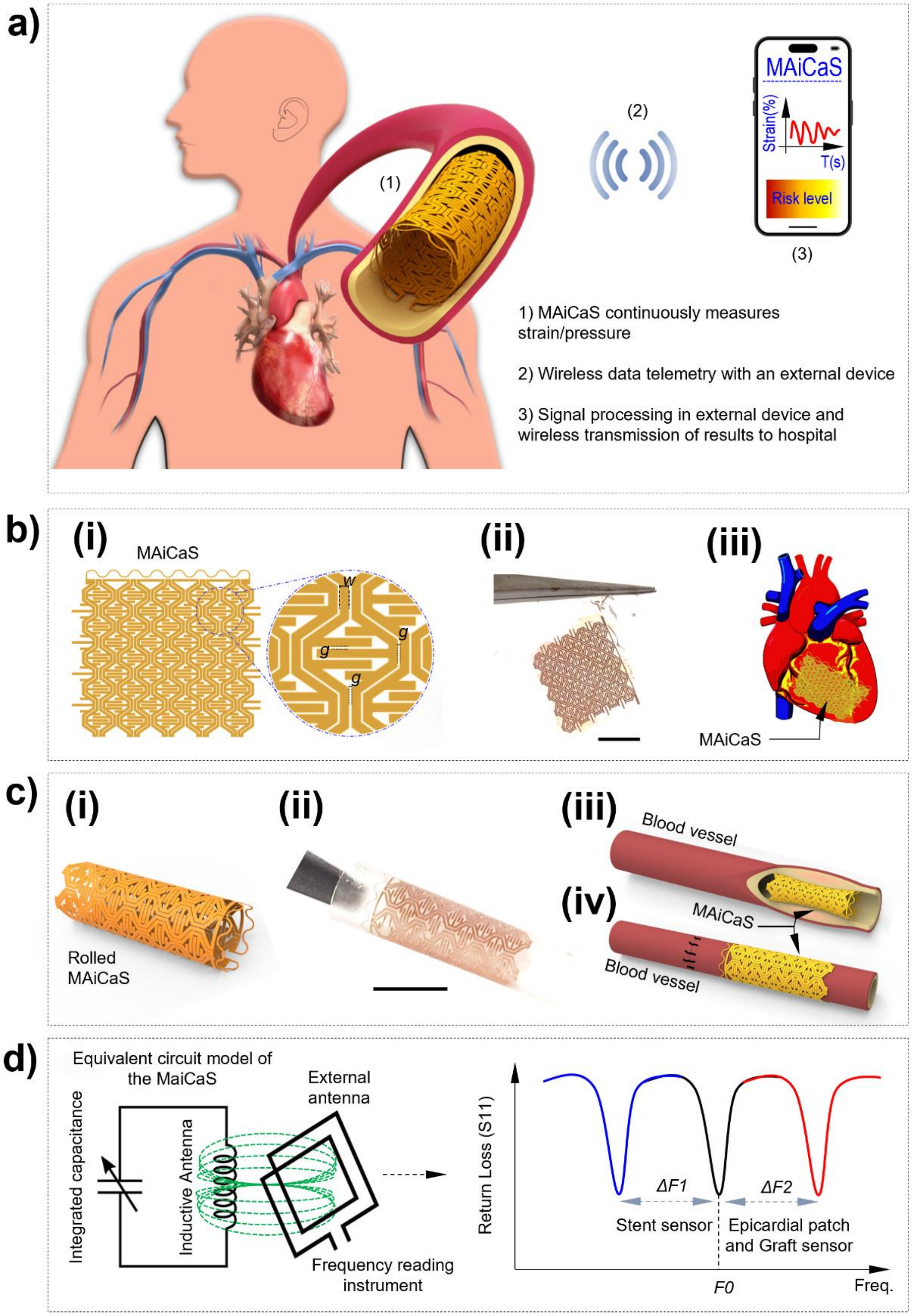

**Figure 1**. Schematic representation and functional concept of the MAiCaS platform for cardiovascular applications. (a) Conceptual illustration of the sensor integrated into a vascular stent, enabling real-time strain measurement and wireless data transmission to a mobile device and remote monitoring via hospital networks. (b) Structural design of the MAiCaS, featuring an interdigitated capacitor and integrated inductor

with an area of 10 mm², metal trace width of w = 120 μm, and inter-electrode gap of g = 30 μm (i); optical image of the fabricated device (scale bar: 0.5 cm) (ii); schematic illustration of MAiCaS operation as an epicardial patch (iii). (c) Rolled configuration of the sensor, demonstrating adaptability for vascular graft and stent applications: schematic illustration (i), optical image of the fabricated MAiCaS in rolled configuration (scale bar: 0.5 cm) (ii), schematic illustration of rolled MAiCaS applied as a stent (iii) and as a vascular graft sensor (iv), with placement inside and around a blood vessel. (d) Equivalent circuit model of the MAiCaS, along with the corresponding resonance frequency shifts associated with its different configurations as an epicardial patch, stent sensor, and vascular graft sensor. When applied as a stent, the resonance frequency of MAiCaS decreases with increasing pressure. In contrast, when configured as an epicardial patch or vascular graft sensor, the resonance frequency increases with increasing strain.

## 2. Results and Discussion

### 2.1. Device structure and operation

The proposed MAiCaS device (Fig. 1), operates as a passive LC circuit, integrating IDE and an inductive antenna to enable wireless strain sensing. The IDEs are strategically configured to manipulate the parasitic capacitance of the inductive antenna, allowing for precise control of capacitance response under mechanical strain. By adjusting electrode spacing, the device can detect subtle arterial deformations, translating them into measurable shifts in resonance frequency. The structural layer of the device was fabricated from a flexible copper layer. The antenna and IDE metallic traces were subsequently patterned by UV laser engraving. Furthermore, the device was encapsulated by a thin (~400 μm) PDMS layer to improve device performance by offering mechanical protection, flexibility, and effective force transmission.

### 2.2. Device fabrication and initial characterization

#### 2.2.1. Fabrication process

As shown in Fig. 2, the cleanroom-free fabrication process began with the preparation of a polydimethylsiloxane (PDMS, Sylgard 184 Silicone Elastomer) base layer with a target thickness of 200 μm on a silicon die as a handling wafer. A PDMS mixture, prepared in an 85:15 ratio, was spin-coated at 4000 rpm for 60 seconds to ensure uniform coating over the silicon substrate. The coated silicon die was then baked at 90°C for 3 minutes to solidify the PDMS layer and prepare it for subsequent processing. Then, a piece of 30 μm-thick copper layer (2 $cm^2$) was thoroughly cleaned and manually laminated over the PDMS layer. To ensure effective adhesion of the film to the PDMS, surface activation was performed by a plasmonic activation for 60 seconds. Following activation, the copper layer (30 μm thick functional layer) was pressed onto the PDMS layer. The layer was then patterned using the UV laser [34], which allowed for precise cutting to create the desired electrode structure. The pattern included IDE electrodes with a minimum spacing of 30 μm and a line width of 120 μm. The overall LC passive sensor dimensions were designed to be 1 $cm^2$. After the patterning process, a second PDMS layer with a thickness of 200 μm was applied for encapsulation. The device was then baked at 90°C for another three minutes, securing the top PDMS layer and creating a stable, integrated LC passive sensor. Once the encapsulation was complete, the sensor structure was peeled off from the silicon die. To assemble the sensor into a cylindrical form (for applications as a stent or graft), two ends were brought together and bonded using a thin PDMS layer at the location of the joint. This final device was cured at 90°C for an additional 3 minutes, resulting in a robust structure for the integrated sensor and inductive antenna. A scanning electron microscope (SEM) image of the fabricated device in the location of the IDE fingers is provided in the Fig. 2j, indicating proper engraving of the copper layer by UV laser. The figure indicates a gap of 30 μm and a width of 120 μm for IDE fingers. Furthermore, confocal microscopy imaging was employed to investigate the thickness of the functional layer. As shown in Fig. 2k, the functional layer has a thickness of ~32 μm.

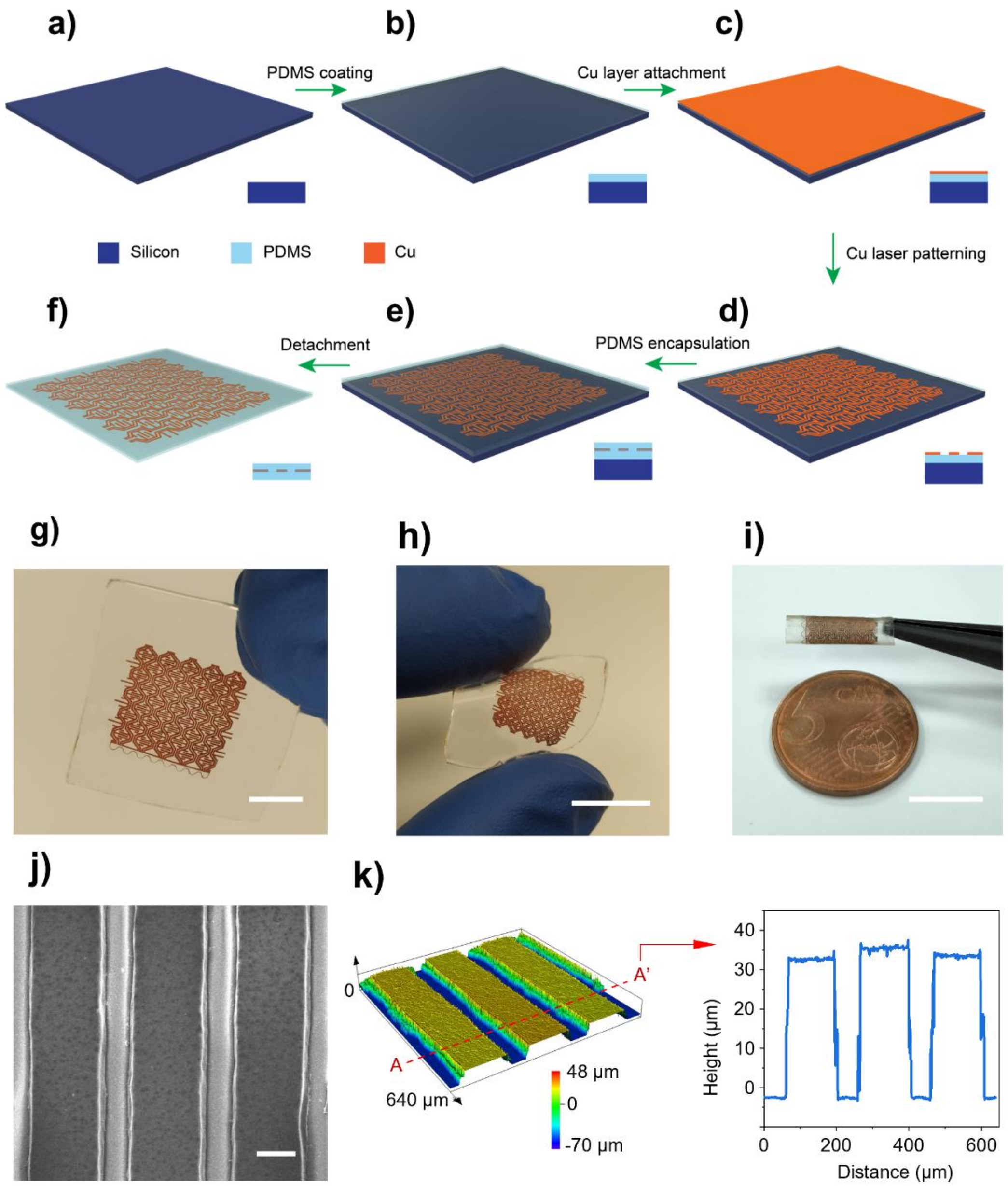

**Figure. 2.** Cleanroom-free fabrication process of the MAiCaS. (a) Bare silicon wafer, (b) Spin-coating of a 200 μm thick PDMS base layer onto a silicon die to form the substrate, followed by plasma activation of the PDMS surface to enhance the adhesion properties during Cu placement, (c) Manual lamination of a 30 μm thick copper film onto the activated PDMS surface, (d) UV laser patterning of the copper layer, defining the sensor structure with a 120 μm copper line width and 30 μm inter-line spacing, (e) Encapsulation of the patterned copper layer with a ~200 μm thick upper PDMS layer, ensuring structural stability and electrical insulation, (f) Peeling off the encapsulated sensor from the silicon die, (g) Optical image of the fully fabricated sensor, showing the precise patterning of the structure (scale bar: 5 mm), (h) Optical image demonstrating the sensor's flexibility (scale bar: 10 mm). (i) Optical image of the sensor in its rolled configuration, illustrating its applicability as a stent or vascular graft sensor (scale bar: 10 mm), (j) SEM

image of the MAiCaS at the location of IDE fingers. The image indicates a 120 μm copper line width and 30 μm inter-line spacing (scale bar: 600 μm), (k) confocal imaging of the MAiCaS at the location of IDE fingers. The plot indicates a height of around 32 μm to the structural layer of the device.

### 2.2.2. Initial characterization:

The initial characterization of the fabricated unrolled MAiCaS was conducted to evaluate its resonance behavior, stability in biological environments, long-term reliability, and wireless interrogation capabilities. The resonance frequency and reflection coefficient ($S_{11}$) of the device were measured to be 1.71 GHz and –14 dB, respectively (measurement was done by an external antenna coupled to a vector network analyzer (VNA)). Then, in order to evaluate the effect of the surrounding environment on the sensor's resonance frequency and $S_{11}$, it was placed over a rat's skin, and measurements indicated a stable resonance behavior for the device (Figure 3a). Further stability evaluations were performed by immersing the sensor in phosphate-buffered saline (PBS) and real human serum, simulating real biological conditions. The measured $S_{11}$ parameter showed a consistent resonance response, indicating minimal impact from biofluid exposure (Figure 3b). To ensure long-term operational stability, the sensor underwent an aging test, where it was stored in air for 16 days at an aging temperature of 70 °C, corresponding to a storage time of 1 year, with periodic $S_{11}$ measurements recorded [34]. The results demonstrated no significant variations in resonance frequency or $S_{11}$ magnitude, confirming its durability and suitability for extended use in biomedical applications (Figure 3c). These findings validate the sensor's robust electrical performance, long-term biostability, and wireless interrogation efficiency in biological environments.

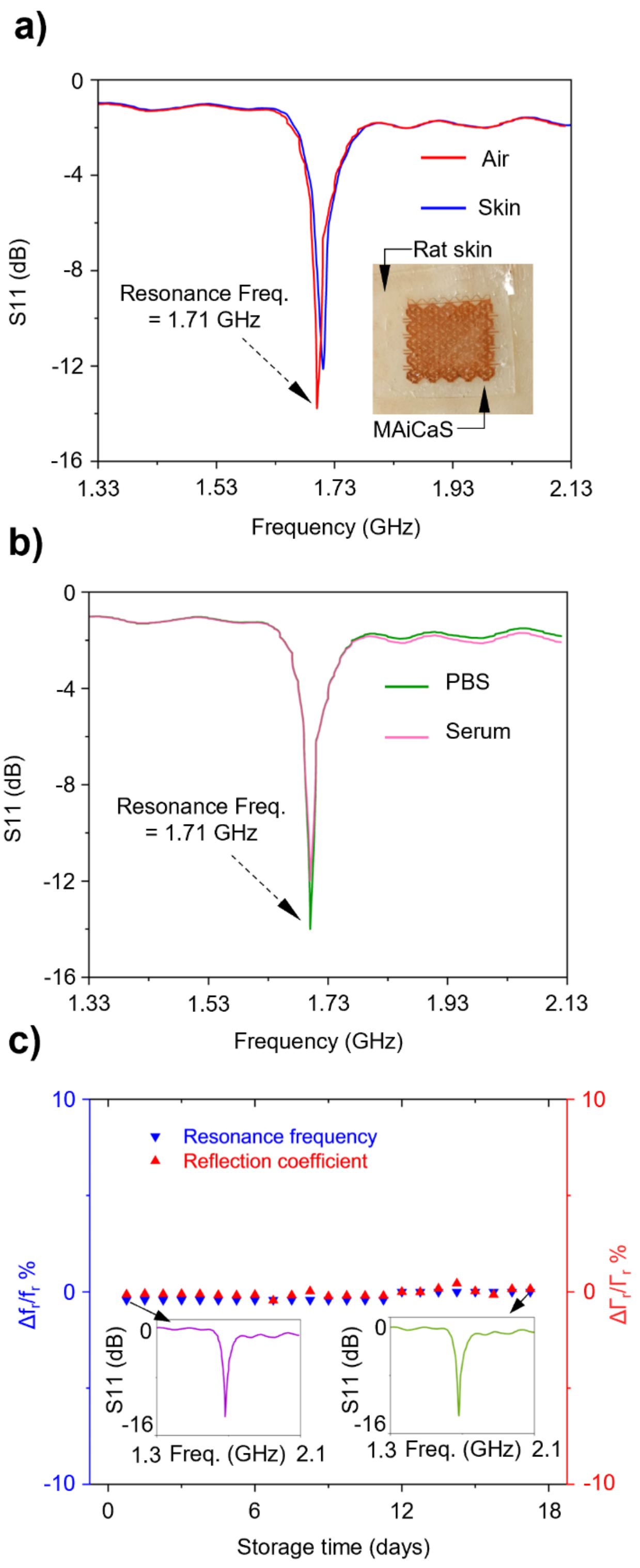

**Figure 3**. Initial characterization of the fabricated unrolled MAiCaS. (a) Measurement of the MAiCaS resonance frequency in air and after placement over a piece of rat skin, (b) Stability evaluation of the sensor when immersed in PBS and real human serum, assessing performance in physiological conditions. (c)

Aging test results indicating the resonance frequency shift and sensor characteristics over a 16-day period where the sensor was maintained at a temperature of 70 °C.

## 2.3. Device characterization

### 2.3.1. Unrolled device characterization for strain sensing and its envisioned application as an epicardial patch for continuous monitoring of left ventricular (LV) strain

The envisioned clinical application of the epicardial patch is continuous monitoring of left ventricular (LV) strain as an early diagnostic tool for arrhythmia and other myocardial dysfunctions. Mechanical signatures often precede or complement electrical abnormalities, and several studies have demonstrated that impaired myocardial strain is strongly associated with arrhythmic risk and adverse cardiac outcomes. Under physiological conditions, the LV typically exhibits longitudinal strain values of 18% to 22% and circumferential strain of 20% to 25%, consistent with normal myocardial contractility [35, 36]. In patients prone to ventricular arrhythmias, global longitudinal strain (GLS) is often reduced to 12% to 15% or lower, accompanied by increased mechanical dispersion (time-to-peak strain differences exceeding ~60 ms), both of which have been identified as predictors of ventricular tachyarrhythmia and sudden cardiac death [26, 37, 38]. By ensuring sensitivity across this clinically relevant strain range, the proposed patch could facilitate real-time detection of abnormal mechanical alternans or regional contractile dysfunction as an early biomarker of arrhythmogenic events. Furthermore, the sensor incorporates a meandered structural layout, which enhances mechanical compliance and allows conformal attachment to the epicardial surface without imposing localized stress, thereby minimizing the risk of tissue irritation or damage during repetitive cardiac motion [39, 40]. The experimental characterization of the unrolled sensor was conducted to evaluate its strain sensitivity and repeatability under controlled mechanical deformation, as illustrated in Fig. 4. The sensor was designed for implantation on soft tissues such as the heart, where it could be used to monitor biomechanical strain variations in real time. To achieve precise deformation, the sensor was mounted on a translation stage using 3D-printed holders that fixed it at both ends while permitting controlled expansion (Fig. 4a). A single-ring external antenna connected to a VNA was employed to measure resonance frequency shifts during strain application. The antenna was positioned beneath the sensor to enable wireless interrogation and record the $S_{11}$ parameter and resonance frequency shifts (Fig. 4a,ii). The translation stage was incrementally displaced from 0 to 2000 μm. Considering the strain relation $\varepsilon = \Delta L/L_0$, where $L_0$ = 10,000 μm is the sensor length, this corresponds to 20% strain. At each step, the $S_{11}$ parameter was recorded to establish the correlation between applied strain and resonance frequency shift. The results, presented in Fig. 4b, demonstrate a clear correlation between applied strain and resonance frequency shift, confirming the sensor's strain sensitivity. The resonance frequency at rest was 1.71 GHz, and upon incremental strain steps of 5%, the resonance frequency shifted to 1.726, 1.74, 1.755, and 1.769 GHz, respectively. The calibration curve, constructed from at least five independent measurements at each strain level, with error bars indicating the standard deviation, is shown in Fig. 4c. The sensor exhibited a sensitivity of ~2.9 MHz per 1% strain. To assess repeatability and durability, the sensor underwent four consecutive expansion–contraction cycles at 5% strain, consistently returning to its baseline state without hysteresis (Fig. 4d). The theoretical formulations confirming the operation of the device are provided in the SI file. In summary, the sensor showed reliable strain sensitivity and cyclic stability within ranges relevant to LV motion, suggesting its suitability for epicardial strain monitoring and possible future use in arrhythmia detection.

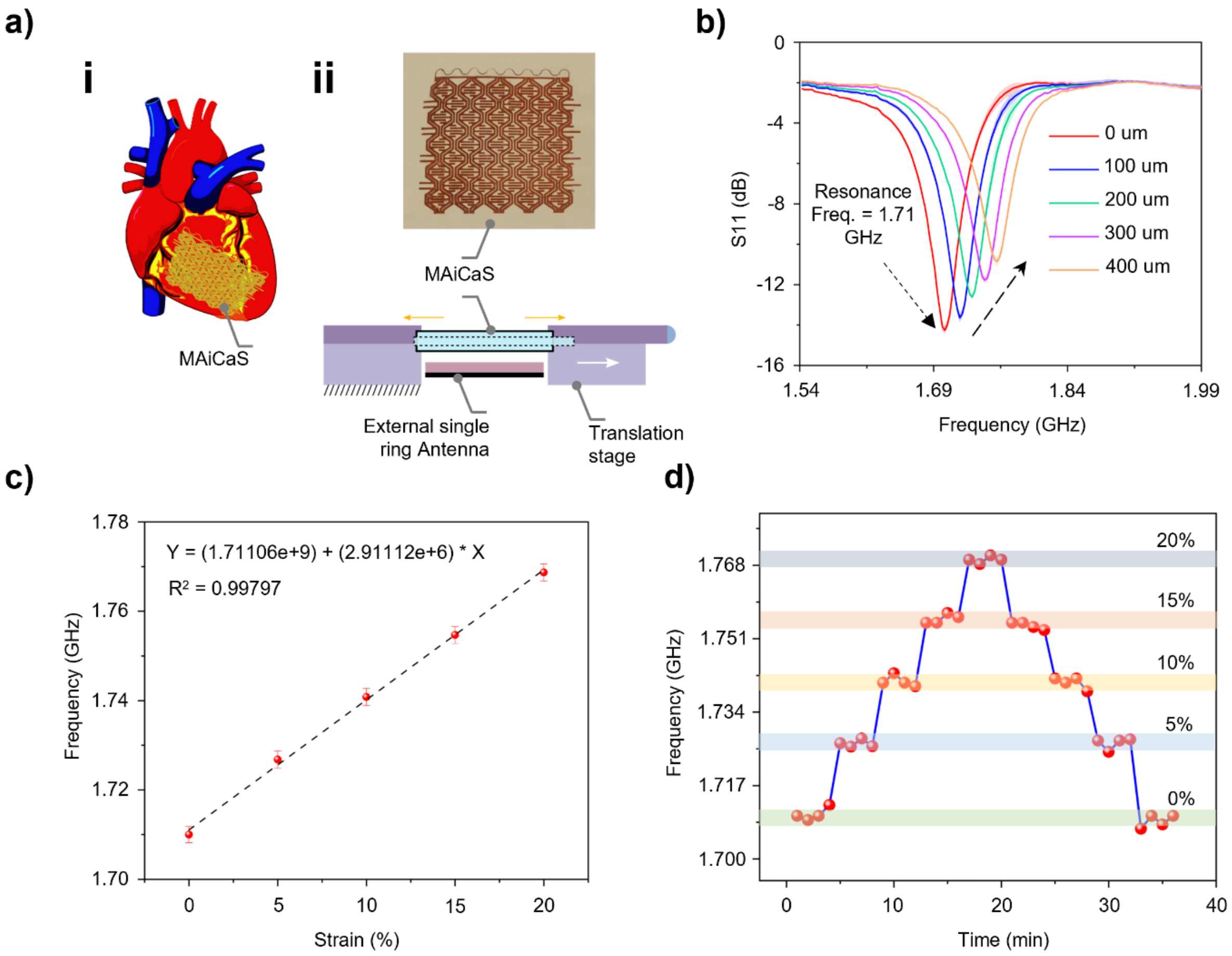

**Figure 4**. Experimental characterization of the unrolled MAiCaS. (a) Schematic and optical representation of the sensor application and experimental setup. (b) Measured resonance frequency shift of the sensor as a function of applied strain via the translation stage, demonstrating its strain sensitivity. (c) Calibration curve of the sensor, with each data point representing the average of at least five measurements, and error bars indicating the standard deviation. (d) Repeatability test confirming the sensor's ability to return to its initial state, validating its elastic behavior and stable performance under cyclic strain application. Each red scatter point denotes the measured resonance frequency during strain application and release.

### 2.3.2. Rolled device characterization for pressure sensing as a possible application for vascular graft-integrated pressure sensor

The envisioned application of the MAiCaS as a vascular graft–integrated pressure sensor addresses several important clinical requirements. In normal physiology, arterial pressures range between ~100–120 mmHg systolic and ~60–80 mmHg diastolic, while pathological conditions such as hypertension, hypotension, or post-surgical complications may extend this range from ~5 to 200 mmHg [41, 42]. Therefore, the sensor must operate reliably across this full pressure range with sufficient resolution (≈1–2 mmHg) to detect early hemodynamic changes. From a diagnostic perspective, continuous pressure monitoring within the graft could enable early detection of stenosis, thrombosis, or endoleaks in aortic repair [43, 44], provide long-term follow-up for congenital defect corrections [45], and improve the management of dialysis access grafts by allowing early intervention before occlusion occurs [46]. To meet these demands, the sensor must combine high sensitivity and stability with mechanical compliance, ensuring that graft function as a load-bearing conduit is not compromised. By fulfilling these requirements, a graft-integrated pressure sensor holds strong potential to advance real-time clinical diagnostics and postoperative monitoring in vascular surgery.

The rolled MAiCaS was evaluated for its response to applied pressure when wrapped around a vessel-like structure, mimicking its potential use as a vascular graft-integrated pressure sensor, as shown in Figure 5. The experimental setup is shown in the SI file, which utilized a water container and a serum hose to generate controlled pressure variations, while a latex hose was used to simulate a biological vessel [47-49]. The sensor was rolled around the latex vessel, forming a cylindrical structure, and was secured in place (Fig. 5a). An external single-ring antenna coupled to a VNA was positioned beneath the setup to facilitate wireless interrogation of the sensor's resonance behavior (Fig. 5a). Pressure was applied by modulating the water container level, which, in turn, induced strain on the MAiCaS by altering the internal pressure within the vessel-like structure [48, 50]. The resonance frequency shift was recorded as a function of applied pressure (measured in mmHg), and the results are plotted in Figure 5b. The response of the MAiCaS for different applied pressures of 50, 100, 150, and 200 are 1.676, 1.698, 1.719, and 1.741 GHz, respectively. A linear correlation between the applied pressure and the sensor's resonance frequency shift was observed, confirming the sensor's high sensitivity to vascular compliance variations. The calibration curve, shown in Figure 5c, was constructed using at least five independent measurements per pressure level, with error bars representing the standard deviation. The MAiCaS exhibited a sensitivity of 0.43 MHz/mmHg, corresponding to a total frequency shift of ≈85.5 MHz across 0–200 mmHg. Considering that clinically relevant hemodynamic monitoring requires a resolution of ~1–2 mmHg [51], this sensitivity is sufficient, as even modest frequency resolution of ≤1 MHz translates to a pressure resolution of <2.5 mmHg. These results confirm that the sensor's sensitivity and dynamic range adequately meet the requirements for continuous pressure monitoring in vascular graft applications.

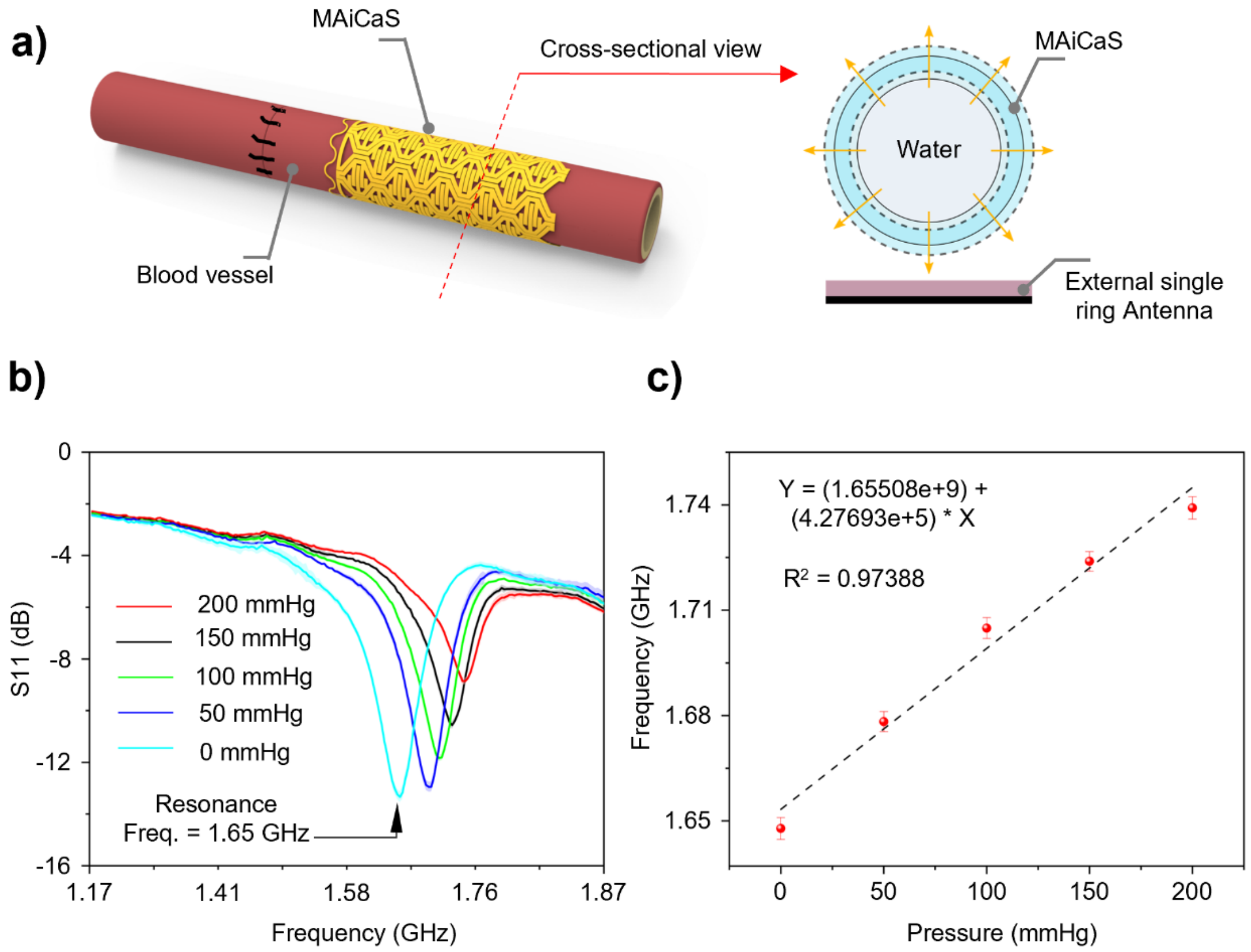

**Figure 5**. Experimental characterization of the rolled MAiCaS in response to applied pressure. (a) Schematic representation of the MAiCaS and experimental setup. (b) Measured resonance frequency shift of the MAiCaS as a function of applied pressure (mmHg), demonstrating its pressure sensitivity. (c) Calibration curve of the sensor, where each data point represents the average of at least five measurements, with error bars indicating the standard deviation.

### 2.3.3. Rolled device characterization for pressure sensing as a possible application for stent-integrated pressure sensor

Another envisioned application of the rolled MAiCaS is as a stent-integrated pressure sensor for intravascular monitoring. Continuous in-stent pressure measurements can provide valuable diagnostic information by revealing early signs of in-stent restenosis, thrombosis, or local hemodynamic changes, which are often undetectable with standard imaging modalities. Physiological intravascular pressures typically range from 60–120 mmHg, while pathological conditions such as hypertension, acute coronary syndromes, or stent malfunction can extend this range between 5 and 200 mmHg [52, 53].

To assess the rolled MAiCaS performance as a stent-integrated deformation sensor, it was inserted inside a serum hose, mimicking its deployment within a vessel, as illustrated in Fig. 6a. A translation stage and custom 3D-printed holders were employed to apply controlled deformation to the sensor. The serum hose acted as a vessel surrogate, holding the sensor inside, while the 3D-printed holders were used to secure the structure. The translation stage was used to compress the sensor radially, simulating in vivo deformation scenarios, such as vessel narrowing or restenosis progression. The reader antenna was positioned beneath the experimental setup, and a VNA was used to record resonance frequency shifts in response to applied deformation (Fig. 6a). The translation stage was gradually moved from 0 to 400 μm, inducing a controlled change in the sensor's diameter. Since this deformation directly altered the rolled sensor's structure, it introduced a corresponding strain variation, which was reflected in the measured resonance frequency shifts. The reflection coefficient was recorded at each deformation step, confirming the sensor's sensitivity to vessel deformation. The developed experimental setup is provided in SI file. The experimental results, shown in Fig. 6b, illustrate a measurable and consistent shift in resonance frequency corresponding to applying a displacement of 100, 200, 300, and 400 μm to the rolled MAiCaS, leading to frequency shifts from 1.53 to 1.55, 1.59, 1.62, and 1.65 GHz, respectively. The calibration curve, obtained from five independent measurements per deformation level, exhibited a highly consistent response with minimal deviation, exhibiting a sensitivity of −309.6kHz/μm, validating the sensor's ability to precisely detect vascular deformation (Fig. 6c). In this characterization, the rolled device was tested under controlled uniaxial displacement loading (0–400 μm) rather than direct intraluminal pressurization. While this does not allow a direct conversion of displacement to luminal pressure, it provides a reproducible way to validate strain tolerance and mechanical robustness under cyclic deformation. Imaging and biomechanical studies report pulsatile circumferential strain in large arteries on the order of ~1–5% per heartbeat (e.g., thoracic aorta controls ≈0.9–3.3%) and show that stent implantation alters local vascular deformation [54, 55]. Consistent with arterial pressure–diameter data, physiologic peak-to-peak diameter changes of a few percent are typical in peripheral and carotid vessels (e.g., ~2.8% femoral, ~6.7% carotid) [56, 57]. For a 3.18 mm lumen, 1–3% cyclic diameter change corresponds to approximately 32–95 μm; thus, the 0–400 μm displacements used here span and exceed physiological deformation, serving as upper-bound mechanical screening of the rolled device for stent-integrated pressure sensing. The findings confirm the sensor's potential for real-time vascular deformation monitoring, making it an effective candidate for detecting restenosis and arterial narrowing.

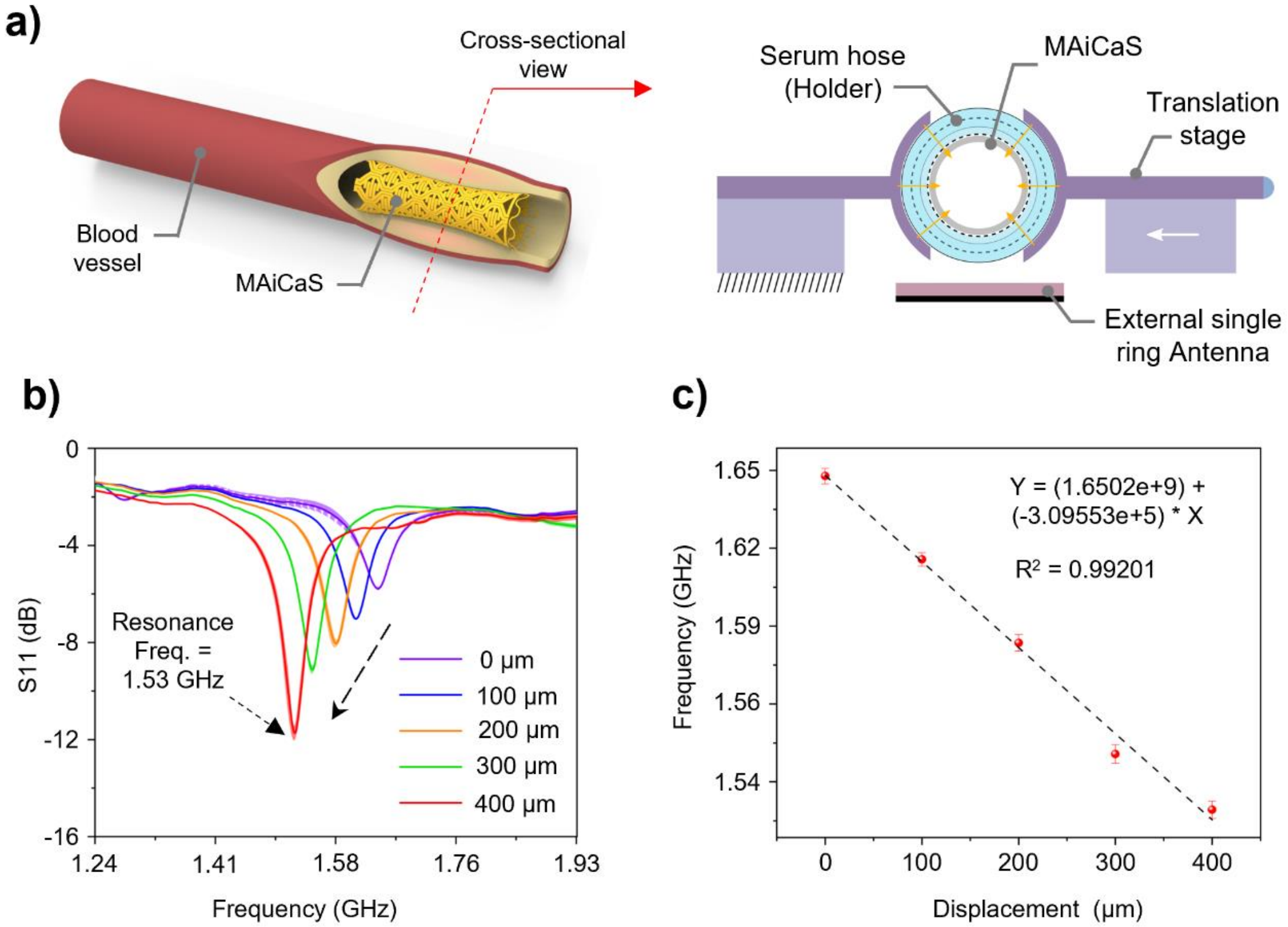

**Figure 6**. Experimental characterization of the rolled MAiCaS in response to applied deformation. (a) Schematic representation of the sensor application and experimental setup. (b) Measured resonance frequency shift of the sensor as a function of applied displacement (μm) using the translation stage, demonstrating its deformation sensitivity. (c) Calibration curve of the sensor, where each data point

represents the average of at least five independent measurements, with error bars indicating the standard deviation.

### 2.3.4. MAiCaS performance validation by on-body human experiments

To further validate the feasibility of the developed MAiCaS for operation on soft, deformable tissues, a preliminary in vivo proof-of-concept experiment was performed on a human finger joint. The unrolled device was conformally attached to the skin surface over the proximal interphalangeal (PIP) joint, and the finger gradually flexed from a neutral position to approximately 90°. This motion-imposed elongation on the sensor substrate (PDMS) as the joint angle increased, thereby mimicking cyclic strain conditions like those encountered in vivo on dynamic biological tissues. Figure 7 summarizes the experimental setup and the resonance response of MAiCaS as a function of PIP joint angle. Figure 7a(i) shows an optical image of MAiCaS secured over the PIP joint. Representative resonance spectra at 0°, 15°, 30°, 60°, and 90° are shown in Fig. 7a(ii–vi); the corresponding resonance frequencies were 1.49, 1.59, 1.70, 1.84, and 1.93 GHz, respectively, exhibiting a monotonic increase with flexion. The angle–frequency trend is summarized in Fig. 7b. Each angle was measured in at least three independent repeats. A linear calibration (Fig. 7c) described the response over 0–90° ($Y = a + b*X$; $a = 1.55161$ GHz, $b = 0.00455$ GHz $deg^{-1}$; $R^2 = 0.9769$), yielding a sensitivity of 4.55 MHz $deg^{-1}$. To assess repeatability and durability, the device was cycled 200 times between 0° and 90°; the resonance frequency returned to baseline after each cycle with negligible hysteresis (Fig. 7d), indicating robust performance under cyclic mechanical loading. These results confirm that MAiCaS can operate continuously on tissue and skin to track finger bending, supporting its suitability for in vivo motion monitoring and motivating broader multi-participant, long-duration studies.

Table 1 compares the structure and main operating parameters of the MAiCaS in this paper with recently published implantable strain sensors with similar applications in the literature. Based on this table, the MAiCaS has advantages in terms of single-step fabrication without needing integration, operational flexibility (three different applications), and operational performance.

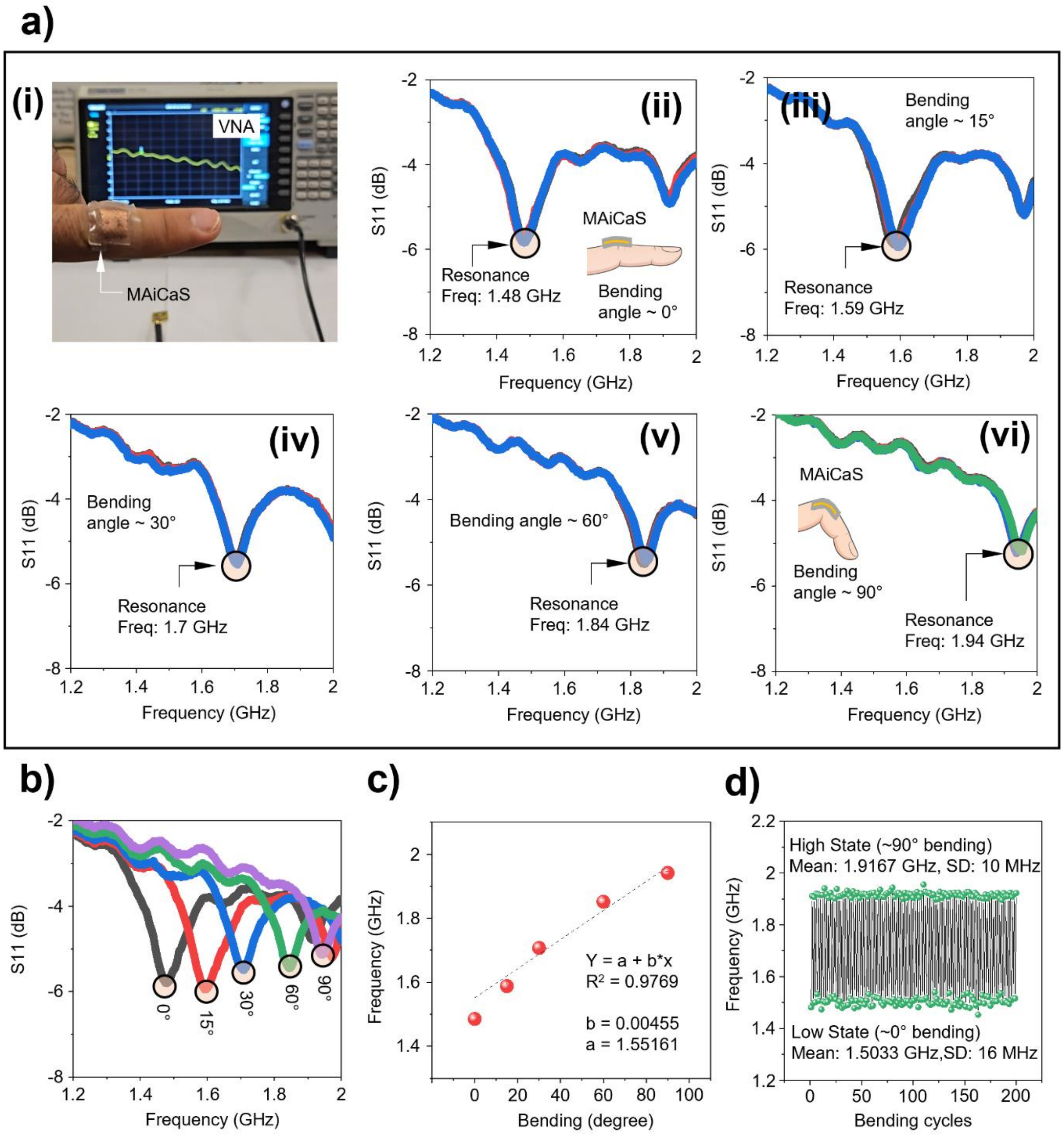

**Figure 7**. MAiCaS performance validation by on-body human experiments. a,i) MAiCaS attachment to the PIP joint for on-body experiments, (ii) resonance frequency of MAiCaS while the bending angle is zero, (iii) 15°, (iv) 30°, (v) 60°, and (vi) 90°, b) resonance frequency of MAiCaS at all bending states, c) extracted calibration curve for on-body experiment indicating the linearity of the MAiCaS response to different bending angles and its repeatability, d) stability test results of the sensor at 0° and 90° of bending of PIP joint for 200 cycles.

**Table 1.** Comparison of the developed integrated sensing platform in this research with previously published sensors.

| Ref. | Type | Fabrication Step | Materials | Cleanroom | Sensor System Integration | Application | Device size (mm) |
|---|---|---|---|---|---|---|---|
| [11] | Stent | • 5 for the stent<br>• 4 for the sensor | • Stainless Steel (SS)<br>• Gold (Au)<br>• Parylene<br>• Polyimide (PI)<br>• Silver Nanoparticles (AgNPs)<br>• PDMS | Yes | Pressure sensor | • Arterial Pressure<br>• Pulse Rate<br>• Flow Rate | Diameter: 5 |
| [2] | Stent | • 5 for the stent<br>• 12 for the sensor | • SU-8 polymer<br>• Chromium (Cr) and Gold (Au)<br>• Titanium (Ti) and Copper (Cu)<br>• PermiNex® 1000, an adhesive<br>• Parylene C | Yes | Pressure sensor | • Cardiovascular pressure | • Diameter: 6<br>• Length: 18 |
| [13] | Stent | • 5 for the stent<br>• 3 for the sensor | • Stainless Steel (SS)<br>• Gold (Au)<br>• Parylene<br>• Polyimide (PI)<br>• Silver Nanoparticles (AgNPs)<br>• Poly(styrene-isoprene-styrene) (SIS)<br>• PDMS | Yes | Strain sensor | • Monitoring of restenosis via arterial stiffness | • Diameter (expanded): 4.8<br>• Length: 25 |
| [14] | Stent | 4 key stages | • SU-8<br>• Titanium (Ti)<br>• Chromium (Cr)<br>• Copper (Cu)<br>• PermiNex | Yes | Dual-pressure sensors are integrated into a single 2D stent structure | • Detecting In-Stent Restenosis<br>• Monitoring Blood Pressure<br>• Monitoring Blood Flow | • Diameter: 3−4<br>• Length: 12 |
| [15] | Stent | • 5 for the stent<br>• 4 for the sensor | • Stainless Steel (SS)<br>• Copper (Cu)<br>• Gold (Au)<br>• Parylene-C<br>• Polyimide (PI)<br>• PDMS | Yes | Sensor is threaded through the stent | • Continuous Monitoring of Stent Edge Restenosis (SER)<br>• Detecting localized hemodynamic pressure changes<br>• Real-time diagnostic capability | • Diameter: 4<br>• Length: 27 |
| [16] | Stent | • 5 key stages | • PDMS<br>• Polyimide<br>• Chromium (Cr) / Gold (Au)<br>• Conductive silver adhesive | Yes | A single device integrating a variable parallel-plate capacitor and a planar spiral inductor to form a passive LC circuit | Blood pressure monitoring | • Diameter: 5<br>• Width: 1.1 |
| [18] | Graft | • 8 key steps | • Polycaprolactone (PCL) for scaffold<br>• Polyvinylidene Fluoride (PVDF) for sensing<br>• Silver Nanowires (AgNWs) for electrodes | Yes | An all-in-one device where a piezoelectric sensing mat is encapsulated between two structural nanofiber mats | Vascular graft for hemodynamics | • Internal diameter: 4<br>• Length: 40 |
| | | | • PDMS for a waterproof layer | | | | |
| [19] | Graft | • 8 key steps | • Polyimide (PI) sheet (source of graphene)<br>• Laser-Induced Graphene (LIG) for sensing<br>• PDMS for substrate and encapsulation<br>• Ag wires for electrical connections | Yes | A seamless integration of a flow sensor into the graft, which is then connected to a separate wireless electronics module. | Vascular graft for hemodynamics | Internal diameter: 2 mm (used for rolling) |
| [20] | Graft | • 5 key steps | • Polyimide for the flexible substrate<br>• Copper/other conductive material for DHA traces<br>• Carbon Black-Polydimethylsiloxane (CB-PDMS) for the sensor<br>• Silicone rubber for encapsulation | Yes | A flexible sensor is directly patterned onto the same polyimide substrate that contains the DHA inductive antenna | Blood flow and pressure monitoring | • split-double helix antenna (DHA) diameter: 3-10<br>• Target vessel diameter: 3-8 |
| [58] | Graft | • 3 key steps | • Lead-free Sodium Niobate (NaNbO3) micro-fibers for the piezoelectric filler<br>• PDMS for the polymer matrix<br>• Gold (Au) for sputtered electrodes<br>• ePTFE (expanded polytetrafluoroethylene) for the vascular graft | Yes | A piezoelectric sensor is bonded onto a vascular graft and connected to a separate inductive coil to form a resonant system. | Monitoring of blood pressure | • Composite surface area: 20 x 50<br>• Graft internal diameter: 6 |
| [59] | Epicardial patch | • 3 key steps | • Molybdenum (Mo) for the bioresorbable conductive mesh<br>• POCO (a citrate-based polymer) for the bioresorbable elastic substrate<br>• PEG-LA-DA hydrogel for the bioadhesive layer | No | Aconductive metal mesh is integrated with a polymer substrate by placing the mesh onto a viscous polymer solution and then curing it | A bioresorbable cardiac patch that provides electrical and mechanical support to promote cardiac tissue regeneration after a myocardial infarction | • Patch size: 30 x 30<br>• Substrate thickness: 0.12<br>• Mesh trace width: 0.1 |
| [60] | Epicardial patch | • 5 key steps | • Silicone elastomer (Ecoflex) for the substrate<br>• Gold (Au) and Titanium (Ti) for interconnects and base electrodes<br>• Polyimide (PI) for electrical insulation<br>• Nanotextured Platinum–Iridium (Pt–Ir) alloy or PEDOT | Yes | Separately fabricated sensors, actuators, and wireless modules integrated onto 3D heart model and encapsulated into a flexible membrane. | • Cardiac electrotherapy to treat arrhythmias<br>• Pacing, defibrillation, and cardiac ablation therapy<br>• Multifunctional platform for spatio-temporal mapping and stimulation | • Substrate thickness: 0.1<br>• Electrode area: 14.5 x 2.5 |
| This work | Multifunctional patch | • Single step | • Silicone elastomer<br>• Copper layer (30 μm thick) | No | No | • Epicardial patch<br>• Graft<br>• Stent integrated sensor | Patch size: 1 $cm^2$ |

## 3. Conclusion

In this work, we developed a passive inductive antenna embedded with capacitive sensors for real-time and continuous monitoring of strain and pressure. The device was designed in a configuration that can be used for three different possible applications, including smart stent, vascular graft, and epicardial patch. The device features a cleanroom-free and single-step fabrication process flow, which eliminates the complexities of the previously demonstrated designs. In vitro experiments validated a strong correlation between geometry modification and electrical response, demonstrating its feasibility for continuous and real-time pressure and strain monitoring. Device quantification was performed by measuring the reflection coefficient in both the unrolled and rolled states. At different configurations of epicardial patch, graft, and stent integrated sensor, the device showed a sensitivity of 2.9 MHz per 1% strain, 0.43 MHz/mmHg, and 309.6kHz/µm, respectively. Furthermore, the operation of MAiCaS was validated through on-body human experiments in an unrolled format, demonstrating repeatable and reliable performance under various bending cycles. The single-step fabrication process, eliminating the need for complex system integration, combined with the device's multifunctional operation, makes it a promising solution for next-generation implantable microdevices for continuous cardiovascular health monitoring.

## 4. Experimental Section

Materials

SYLGARD 184 silicone elastomer kit was purchased from Dow Corning, USA. Flexible copper layer was purchased from DuPont, USA.

Equipment

UV Laser Prototyping System (LPKF ProtoLaser U4), vector network analyzer (SVA 1032X, Siglent Technologies, Germany), Plasma cleaning (Harrick Plasma, PDC-002HPCE), PDMS Mixer (Speedmixer DAC 150.1 FVZ), Stereo Microscope (Zeiss Stemi 305), confocal laser microscope LEXT OLS5000 (Olympus, Japan).

## Statistical analysis

The mean ± s.d. values were used to report the results. OriginLab software (v.9.65) was used to conduct statistical analyses. Each condition was tested in at least three replicates for all experiments.

## Acknowledgements

H.M. acknowledges the support through a Marie Skłodowska-Curie Individual Postdoctoral Fellowship (H2020-MSCA-IF- 2021-101068646, HAMP). The authors acknowledge the use of the services and facilities of Koç University Surface Science and Technology Center (KUYTAM) and n2STAR-Koç University Nanofabrication and Nanocharacterization Center for Scientific and Technological Advanced Research.

## Conflict of Interest

The authors declare no conflict of interest.